\documentclass[
 aip,
 jcp,
 amsmath,
 amssymb,
]{revtex4-1}
\usepackage[utf8]{inputenc}
\usepackage{array}
\usepackage{amssymb}
\usepackage{amsmath}
\usepackage{float}
\usepackage{xcolor}
\usepackage{graphicx}
\usepackage{braket}

\begin{document}

\title{Exact closed-form expression for unitary spin-adapted fermionic singlet double excitation operators}
\author{Erik Rosendahl Kjellgren}
\email{kjellgren@sdu.dk}
\affiliation{Department of Physics, Chemistry and Pharmacy,
University of Southern Denmark, Campusvej 55, 5230 Odense, Denmark.}
\author{Karl Michael Ziems}
\affiliation{School of Chemistry, University of Southampton, Highfield, Southampton SO17 1BJ, United Kingdom}
\affiliation{Department of Chemistry, Technical University of Denmark, Kemitorvet Building 207, DK-2800 Kongens Lyngby, Denmark.}
\author{Peter Reinholdt}
\affiliation{Department of Physics, Chemistry and Pharmacy,
University of Southern Denmark, Campusvej 55, 5230 Odense, Denmark.}
\author{Stephan P. A. Sauer}
\affiliation{Department of Chemistry, University of Copenhagen, DK-2100 Copenhagen \O.}
\author{Sonia Coriani}
\affiliation{Department of Chemistry, Technical University of Denmark, Kemitorvet Building 207, DK-2800 Kongens Lyngby, Denmark.}
\author{Jacob Kongsted}
\affiliation{Department of Physics, Chemistry and Pharmacy,
University of Southern Denmark, Campusvej 55, 5230 Odense, Denmark.}

\begin{abstract}
We derive exact closed-form expressions for the matrix exponential of the anti-Hermitian spin-adapted singlet double excitation fermionic operators.
These expressions enable the efficient implementation of such operators within unitary product state frameworks targeting conventional hardware, and allow for the implementation of ansatze that guarantee convergence to specific spin symmetries.
Moreover, these exact closed-form expressions might also lay the groundwork for constructing spin-adapted circuits for quantum devices.
\end{abstract}


\maketitle

\section{Introduction}

With the advent of quantum computing for the use in computational quantum chemistry, the interest in unitary parameterizations for electronic wave functions has risen.
Among these parameterizations are the hardware-efficient ans{\"a}tze\cite{Kandala2017-fa}, qubit ans{\"a}tze\cite{Ryabinkin2018-jy,Tang2021-gc}, qubit-excitation-based ans{\"a}tze\cite{Yordanov2021-ws} and fermionic excitation-based ans{\"a}tze\cite{Grimsley2019-ix,Lee2019-hj,Burton2024-er,Burton2023-bq,Anselmetti2021-su}.
The fermionic excitation-based ans{\"a}tze are among the most chemically motivated ones, since they conserve the number of $\alpha$ and $\beta$ electrons, ensure the anti-symmetry of the fermionic wave function, and produce an eigenfunction of the z-component of the total spin $\hat{S}_z$.
These ans{\"a}tze are based on 
generic
fermionic excitation operators,
\begin{align}
    \hat{T}_{i_\sigma}^{a_\sigma}&=\hat{a}_{a_\sigma}^\dagger\hat{a}_{i_\sigma}\label{eq:T_single}\\
    \hat{T}_{i_\sigma j_{\tau}}^{a_\sigma b_\tau}&=\hat{a}_{a_\sigma}^\dagger\hat{a}_{b_\tau}^\dagger\hat{a}_{j_\tau}\hat{a}_{i_\sigma}\label{eq:T_double}\\
    &...\label{eq:T_...}
\end{align}
with the indices $i,j,a,b$ referring to spatial orbital indices, and $\sigma,\tau$ to the electron spin.
The anti-Hermitian form of the fermionic operator is given as
\begin{equation}
    \hat{G}_J=\hat{T}_J-\hat{T}_J^\dagger\label{eq:antiH_form}
\end{equation}
with $J$ being a compound index for the indices used in Eqs. (\ref{eq:T_single})--(\ref{eq:T_...}).
So, for instance $\hat{G}^{a\sigma}_{i\sigma} = \hat{T}^{a\sigma}_{i\sigma} - \hat{T}^{i\sigma\dagger}_{a\sigma}$.
The anti-Hermitian form of the fermionic excitation operators gives rise to a unitary parameterization through a matrix exponentiation,
\begin{equation}
    \boldsymbol{U}(\theta_J)=\exp\left(\theta_J\hat{G}_J\right)
\end{equation}
where $\theta_J$ is a free parameter.
This anti-Hermitian form in Eq. (\ref{eq:antiH_form}) fulfills the polynomial relation
\begin{equation}
    \hat{G}_J=-\hat{G}_J^3~,
\end{equation}
this relationship can be used to write the matrix exponential $\exp\left(\theta\hat{G}_J\right)$ in a closed form,
\begin{equation}
    \exp\left(\theta\hat{G}_J\right) = \hat{I} + \hat{G}_J\sin(\theta) + \hat{G}_J^2(1-\cos(\theta))~.
\label{eq:simple_fermionic_closed_form}
\end{equation}
The identity in Eq.~(\ref{eq:simple_fermionic_closed_form}) has been used in previous works\cite{Kottmann2021-dh,Chen2021-sc,Rubin2021-ie,Filip2020-zi,Evangelista2019-le} as well as in an alternative formulation that performs a transformation of a fermionic operator\cite{Evangelista2024-iz}.
Utilizing the closed form in Eq. (\ref{eq:simple_fermionic_closed_form}) enables efficient implementation of the unitary form of these fermionic operators and is utilized in codes such as TenCirChem\cite{Li2023-dp} and SlowQuant\cite{slowquant}, as well as in recent works aiming for 
high-performance\cite{Mullinax2025-cw,Mullinax2024-bg} using an equivalent form\cite{Chen2021-sc}.
In addition, it also enables efficient derivatives through the adjoint differentiation method\cite{Jones2020-up}.
For fermionic operators, exact efficient quantum circuits are also known\cite{Yordanov2020-an,Anselmetti2021-su,Magoulas2023-wi}.
A shortcoming of using generic fermionic operators to parametrize the wave function is that they are not guaranteed to produce wave functions that are eigenfunctions of the total spin angular momentum operator $\hat{S}^2$.
However, operators that produce wave functions that are eigenfunctions of $\hat{S}^2$ can be constructed and are called spin-adapted operators.
Utilizing spin-adapted operators for the wave function parameterization is especially important for multi-state methods such as SS-VQE\cite{Nakanishi2019-qf}, MC-VQE\cite{Parrish2019-oe}, MORE-ADAPT-VQE\cite{Grimsley2025-su}, variance-VQE\cite{Zhang2022-qw}, and SA-VQE\cite{yalouz2021state} since this allows targeting only states of a specific spin symmetry.

The spin-adapted (SA) singlet single excitation operator can be written as,
\begin{equation}
    ^\text{SA}\hat{T}_{ai} = \frac{1}{\sqrt{2}}\hat{E}_{ai}
\end{equation}
with $\hat{E}_{ai}=\hat{a}^\dagger_{a_\alpha}\hat{a}_{i_\alpha}+\hat{a}^\dagger_{a_\beta}\hat{a}_{i_\beta} \equiv (\hat{T}^{a_\alpha}_{i_{\alpha}} + \hat{T}^{a_\beta}_{i_{\beta}})$.
The corresponding anti-Hermitian spin-adapted singlet single excitation operator $^\text{SA}\hat{G}_{ai}$ 
takes the following form: 
\begin{equation}
    ^\text{SA}\hat{G}_{ai}={}^\text{SA}\hat{T}_{ai} - {}^\text{SA}\hat{T}_{ai}^\dagger=\frac{1}{\sqrt{2}}\left(\hat{G}_{i_\alpha}^{a_\alpha}+\hat{G}_{i_\beta}^{a_\beta}\right)
\end{equation}
Using $\left[\hat{G}_{i_\alpha}^{a_\alpha},\hat{G}_{i_\beta}^{a_\beta}\right]=0$, allows us to write the matrix exponential in the form,
\begin{align}
    \exp\left(\theta\ {}^\text{SA}\hat{G}_{ai}\right)&=
    \exp\left(\frac{\theta}{\sqrt{2}}\hat{G}_{i_\alpha}^{a_\alpha}\right)\exp\left(\frac{\theta}{\sqrt{2}}\hat{G}_{i_\beta}^{a_\beta}\right)\\
    &= \left\{\hat{I} + \hat{G}_{i_\alpha}^{a_\alpha}\sin\left(\frac{\theta}{\sqrt{2}}\right) + \hat{G}_{i_\alpha}^{a_\alpha 2}\left(1-\cos\left(\frac{\theta}{\sqrt{2}}\right)\right)\right\}\nonumber\\
    &\quad\times\left\{\hat{I} + \hat{G}_{i_\beta}^{a_\beta}\sin\left(\frac{\theta}{\sqrt{2}}\right) + \hat{G}_{i_\beta}^{a_\beta 2}\left(1-\cos\left(\frac{\theta}{\sqrt{2}}\right)\right)\right\}\label{eq:sa_single_simple_singles}
\end{align}
Due to the form of Eq. (\ref{eq:sa_single_simple_singles}), the spin-adapted singlet single excitation operators can utilize known circuits from the fermionic singlet single excitation operators\cite{Yordanov2020-an} and lead to a parameter reduction compared to the generic form. 

However, contrary to the spin-adapted anti-Hermitian singlet single fermionic excitation operators, the spin-adapted anti-Hermitian singlet double fermionic excitation operators do not decompose into commuting generic fermionic anti-Hermitian operators. 
Hence, a closed-form expression for the matrix exponential of the spin-adapted singlet double excitation operators does not take a simple product similar to the matrix exponential of the spin-adapted singlet single excitation operator in  Eq.~(\ref{eq:sa_single_simple_singles}), and an expression for the matrix exponential of the spin-adapted singlet double excitation operators is unknown to date. 
In the following section, we present exact closed-form expressions for the spin-adapted singlet double fermionic excitation operators.
While completing this work, we became aware of the work of Magoulas and Evangelista\cite{Magoulas2025}, who derived the closed-form expression for the matrix exponential of the anti-
Hermitian form of spin-adapted singlet double fermionic excitations.

\section{Spin-adapted fermionic doubles}

The spin-adapted singlet double fermionic excitation operators take the form,\cite{Paldus1977,Piecuch1989,Packer1996}
\begin{align}
    {}^\text{SA}\hat{T}_{aibj} &= \frac{1}{2\sqrt{\left(1+\delta_{ab}\right)\left(1+\delta_{ij}\right)}}\left(\hat{E}_{ai}\hat{E}_{bj} + \hat{E}_{aj}\hat{E}_{bi}\right)\label{eq:T_intermediate_singlet}\\
    {}^\text{SA}\hat{T}^{\prime}_{aibj} &= \frac{1}{2\sqrt{3}}\left(\hat{E}_{ai}\hat{E}_{bj} - \hat{E}_{aj}\hat{E}_{bi}\right)\label{eq:T_intermediate_triplet}
\end{align}
with Eq. (\ref{eq:T_intermediate_singlet}) stemming from the intermediate singlet single connection to the singlet double excitation, and Eq. (\ref{eq:T_intermediate_triplet}) from the intermediate triplet single connection to the singlet double excitation.
This gives five different cases, since $\hat{T}^{\prime}_{aibj}$ is zero if any of the indices are equal,
\begin{align}
    {}^\text{SA}\hat{T}_{aiai} &= \frac{1}{2}\hat{E}_{ai}\hat{E}_{ai}\label{eq:sa_case1}\\
    {}^\text{SA}\hat{T}_{aiaj} &= \frac{1}{2\sqrt{2}}\left(\hat{E}_{ai}\hat{E}_{aj} + \hat{E}_{aj}\hat{E}_{ai}\right)\label{eq:sa_case2}\\
    {}^\text{SA}\hat{T}_{aibi} &= \frac{1}{2\sqrt{2}}\left(\hat{E}_{ai}\hat{E}_{bi} + \hat{E}_{ai}\hat{E}_{bi}\right)\label{eq:sa_case3}\\
    {}^\text{SA}\hat{T}_{aibj} &= \frac{1}{2}\left(\hat{E}_{ai}\hat{E}_{bj} + \hat{E}_{aj}\hat{E}_{bi}\right)\label{eq:sa_case4}\\
    ^\text{SA}\hat{T}^{\prime}_{aibj} &= \frac{1}{2\sqrt{3}}\left(\hat{E}_{ai}\hat{E}_{bj} - \hat{E}_{aj}\hat{E}_{bi}\right)\label{eq:sa_prime_case1}
\end{align}
Here, we implied that $a\neq b$ and $i\neq j$. In the following, we derive the closed form of the anti-Hermitian versions of Eqs.~(\ref{eq:sa_case1})--(\ref{eq:sa_prime_case1}) .

\subsection{Case ${}^{\mathrm{SA}}\hat{G}_{aiai}$}

This case reduces to a fermionic double excitation operator since $\hat{E}_{ai}\hat{E}_{ai}=2\hat{T}^{a_\beta a_\alpha}_{i_\beta i_\alpha}$, yielding the anti-Hermitian form of 
Eq. (\ref{eq:sa_case1}) as
\begin{equation}
    ^\text{SA}\hat{G}_{aiai} = \hat{G}^{a_\alpha a_\beta}_{i_\alpha i_\beta}
\end{equation}
This already has the known exact closed-form discussed above 
(Eq. (\ref{eq:simple_fermionic_closed_form})),
\begin{equation}
    \exp\left(\theta\ ^\text{SA}\hat{G}_{aiai}\right) = \hat{I} + \ ^\text{SA}\hat{G}_{aiai}\sin(\theta) + \ ^\text{SA}\hat{G}_{aiai}^2(1-\cos(\theta))~.
\end{equation}
Thus, this specific case of spin-adapted singlet double excitations can also utilize known circuits for fermionic operators\cite{Yordanov2020-an}, and is utilized in spin-adapted ansatze using singles and pair-doubles\cite{Anselmetti2021-su,Burton2024-er,Lee2019-hj,Burton2023-bq}. 
For the remaining cases, Eqs. (\ref{eq:sa_case2})--(\ref{eq:sa_prime_case1}), no such reduction can be made.

\subsection{Case ${}^{\mathrm{SA}}\hat{G}_{aiaj}$ and ${}^{\mathrm{SA}}\hat{G}_{aibi}$}
These two cases are the anti-Hermitian versions of the spin-adapted fermionic double excitations operators in Eq. (\ref{eq:sa_case2}) and (\ref{eq:sa_case3}) where two indices match.
They can be shown to share the following polynomial relation for powers of the excitation operator,
\begin{equation}
    ^\text{SA}\hat{G}_{aiaj}^5 = A\ ^\text{SA}\hat{G}_{aiaj}+B\ ^\text{SA}\hat{G}_{aiaj}^3\label{eq:G_poly_5}
\end{equation}
with $A=-\frac{1}{2}$ and $B=-\frac{3}{2}$.
The coefficients $A$ and $B$ were determined by minimization with SLSQP\cite{kraft1988software} using SciPy\cite{2020SciPy-NMeth} together with SlowQuant\cite{slowquant} and SymPy\cite{10.7717/peerj-cs.103} as a symbolic backend.
As the polynomial relation is identical for both ${}^{\mathrm{SA}}\hat{G}_{aiaj}$ and ${}^{\mathrm{SA}}\hat{G}_{aibi}$, the derivation of the closed-form expression will be made only considering ${}^{\mathrm{SA}}\hat{G}_{aiaj}$.

The exponential of an excitation operator can be expanded as
\begin{equation}
    \exp\left(\theta\ {}^{\mathrm{SA}}\hat{G}_{aiaj}\right) = \hat{I} + \sum_{k=1}^\infty\frac{\theta^k}{k!}\ {}^{\mathrm{SA}}\hat{G}_{aiaj}^k~.
\end{equation}
Given that Eq. (\ref{eq:G_poly_5}) gives an expression for the fifth-order term as a combination of first and third-order terms, one can  recast the matrix exponential in a polynomial form that truncates at the fourth power of $^\text{SA}\hat{G}_{aiaj}$,
\begin{equation}
    \exp\left(\theta\ ^\text{SA}\hat{G}_{aiaj}\right) = \hat{I} + \sum_{n=1}^4 f_n\left(\theta\right)\ ^\text{SA}\hat{G}_{aiaj}^n~. \label{eq:gen_fk}
\end{equation}
where the expansion coefficients $f_n$ are polynomial functions of the parameter $\theta$.
The explicit expression of $f_n\left(\theta\right)$ must be determined and arises from the combination of the $\frac{\theta^k}{k!}$, $A$, and $B$ factors.
Using Eq. (\ref{eq:G_poly_5}), let us consider the decomposition of the first and third functions in Eq. (\ref{eq:gen_fk}),
\begin{align}
    f_1(\theta)&=\theta  + \frac{\theta^5}{5!}A + \frac{\theta^7}{7!}AB + \frac{\theta^9}{9!}(A^2+AB^2) + ...\label{eq:f1_explicit}\\
    f_3(\theta)&=\frac{\theta^3}{3!}  + \frac{\theta^5}{5!}B + \frac{\theta^7}{7!}(A+B^2) + \frac{\theta^9}{9!}(2AB+B^3) + ...
\end{align}
It can be seen that the coefficients of the $\frac{\theta^k}{k!}$ terms are
\begin{align}
    K_{1}^{[1]}&=A,\quad K_{2}^{[1]}=AB,\quad K_{3}^{[1]}=A^2+AB^2, \quad\label{eq:k1_example} ...\\
    K_{1}^{[3]}&=B,\quad K_{2}^{[3]}=A +B^2,\quad K_{3}^{[3]}=2AB+B^3, \quad\label{eq:k3_example} ...
\end{align}
Here, the superscript in square parentheses indicates which function the coefficients are associated with, that is $K^{[1]}_n$ is associated with $f_1$ and $K^{[3]}_n$ is associated with $f_3$.
By combining Eqs. (\ref{eq:f1_explicit})--(\ref{eq:k3_example}), one obtains the following expressions for the functions,
\begin{align}
    f_1(\theta)&=\theta  + \sum_{n=1}^\infty\frac{\theta^{2n+3}}{(2n+3)!}K^{[1]}_n\label{eq:f1_tmp_form}\\
    f_3(\theta)&=\frac{\theta^3}{3!}  + \sum_{n=1}^\infty\frac{\theta^{2n+3}}{(2n+3)!}K^{[3]}_n\label{eq:f3_tmp_form}
\end{align}
The coefficients in Eq. (\ref{eq:k1_example}) and Eq. (\ref{eq:k3_example}) can be seen to follow a recurrence relation,
\begin{align}
    K_{n}^{[1]} &= AK_{n-1}^{[3]}\\
    K^{[3]}_n &= BK^{[3]}_{n-1}+ K^{[1]}_{n-1}
\end{align}
It can be noted that the recurrence relations are linear, and a closed-form solution can be found from the eigenvalues of the matrix equation form of the recurrence relations,
\begin{equation}
    \begin{bmatrix}
K^{[1]}_n \\
K^{[3]}_n
\end{bmatrix}
=
\begin{bmatrix}
0 & A \\
1 & B
\end{bmatrix}
\begin{bmatrix}
K^{[1]}_{n-1} \\
K^{[3]}_{n-1}
\end{bmatrix}.\label{eq:recurence_matrix_form}
\end{equation}
We obtain the following solution,
\begin{align}
    K^{[1]}_n &= A(c_1\lambda_1^{n-1}+c_2\lambda_2^{n-1})\label{eq:K1_solution}\\
    K^{[3]}_n &= c_1\lambda_1^{n}+c_2\lambda_2^{n}\label{eq:K3_solution}
\end{align}
with $\lambda_1$ and $\lambda_2$ being the eigenvalues of the coefficient matrix in Eq. (\ref{eq:recurence_matrix_form}), and $c_1$ and $c_2$ being coefficients determined from the initial condition $K^{[1]}_1=A$ and $K^{[3]}_1=B$, found by solving,
\begin{equation}
    \begin{bmatrix}
A \\
B
\end{bmatrix}
=
\begin{bmatrix}
A\lambda_1^{0} & A\lambda_2^0 \\
\lambda_1^1 & \lambda_2^1
\end{bmatrix}
\begin{bmatrix}
c_1 \\
c_2
\end{bmatrix} \rightarrow \begin{bmatrix}
c_1 \\
c_2
\end{bmatrix} = \begin{bmatrix}
A\lambda_1^{0} & A\lambda_2^0 \\
\lambda_1^1 & \lambda_2^1
\end{bmatrix}^{-1}    \begin{bmatrix}
A \\
B
\end{bmatrix}
\end{equation}
The eigenvalues and coefficients are found to be $\lambda_1=\frac{B-\sqrt{4A+B^2}}{2}$, $\lambda_2=\frac{B+\sqrt{4A+B^2}}{2}$, $c_1=\frac{1}{2}-\frac{B}{2\sqrt{4A+B^2}}$ and $c_1=\frac{1}{2}+\frac{B}{2\sqrt{4A+B^2}}$.
By inserting Eq. (\ref{eq:K1_solution}) into Eq.~(\ref{eq:f1_tmp_form}) and Eq. (\ref{eq:K3_solution}) into 
Eq.~(\ref{eq:f3_tmp_form}), the functions now take the following form,
\begin{align}
    f_1(\theta)&=\theta  + \sum_{n=1}^\infty\frac{\theta^{2n+3}}{(2n+3)!}\sum_{i=1}^2Ac_i\lambda_i^{n-1}\label{eq:f1_inf_poly}\\
    f_3(\theta)&=\frac{\theta^3}{3!}  + \sum_{n=1}^\infty\frac{\theta^{2n+3}}{(2n+3)!}\sum_{i=1}^2c_i\lambda_i^{n}\label{eq:f3_inf_poly}
\end{align}
One can identify these sums to take the closed form,
\begin{equation}
    \sum_{n=1}^\infty\frac{\theta^{2n+3}}{(2n+3)!}\lambda_i^{n-m} = \frac{1}{\lambda_i^{3/2+m}}\left(\sinh\left(\theta\sqrt{\lambda_i}\right)-\theta\sqrt{\lambda_i}-\frac{\left(\theta\sqrt{\lambda_i}\right)^3}{3!}\right)\label{eq:coeffs_closed_form}
\end{equation}
with $m=1$ for the $f_1$ case and $m=0$ for the $f_3$ case.
Inserting Eq. (\ref{eq:coeffs_closed_form}) into Eqs. (\ref{eq:f1_inf_poly}) and (\ref{eq:f3_inf_poly}) yields,
\begin{align}
    f_1(\theta)&=\theta  + \sum_{i=1}^2\frac{Ac_i}{\lambda_i^{5/2}}\left(\sinh\left(\theta\sqrt{\lambda_i}\right)-\theta\sqrt{\lambda_i}-\frac{\left(\theta\sqrt{\lambda_i}\right)^3}{3!}\right)\\
    f_3(\theta)&=\frac{\theta^3}{3!}  + \sum_{i=1}^2\frac{c_i}{\lambda_i^{3/2}}\left(\sinh\left(\theta\sqrt{\lambda_i}\right)-\theta\sqrt{\lambda_i}-\frac{\left(\theta\sqrt{\lambda_i}\right)^3}{3!}\right)
\end{align}
Since the eigenvalues $\lambda_i$ are all negative, one can use that $\sinh(\mathrm{i}x)=\mathrm{i}\sin(x)$. Further inserting 
$\lambda_1=\frac{B-\sqrt{4A+B^2}}{2}$, $\lambda_2=\frac{B+\sqrt{4A+B^2}}{2}$, $c_1=\frac{1}{2}-\frac{B}{2\sqrt{4A+B^2}}$, $c_2=\frac{1}{2}+\frac{B}{2\sqrt{4A+B^2}}$, $A=-\frac{1}{2}$, and $B=-\frac{3}{2}$, these expressions reduce to,
\begin{align}
    f_1(\theta)&=\sum_{i=1}^2k_i^{(1)}\sin\left(\theta S_i\right)\\
    f_3(\theta)&=\sum_{i=1}^2k_i^{(3)}\sin\left(\theta S_i\right)
\end{align}
with real coefficients $S_i$ and $k_i^{(n)}$ given in Table \ref{tab:case_aiaj}. 
As it can be noted, all powers in $\theta$ have canceled out.
\begin{table}[H]
    \centering
\caption{Values of coefficients in Eq. (\ref{eq:case_aiaj}).}
\begin{tabular}{ c | c | c | c | c | c }
 $n$ & $k_n^{(1)}$ & $k_n^{(3)}$ & $k_n^{(2)}$ & $k_n^{(4)}$ & $S_n$\\
 \hline
 $1$ & $-1$ & $-2$ & $1$ & $2$ & $1$ \\
 $2$ & $2\sqrt{2}$ & $2\sqrt{2}$ & $-4$ & $-4$ & $\frac{\sqrt{2}}{2}$
 \\\hline
\end{tabular}
\label{tab:case_aiaj}
\end{table}
Carrying out a similar derivation for the even terms in Eq.~(\ref{eq:gen_fk}), we find
\begin{align}
    f_2(\theta)&=\sum_{i=1}^2k_i^{(2)}(\cos\left(\theta S_i\right)-1)\\
    f_4(\theta)&=\sum_{i=1}^2k_i^{(4)}(\cos\left(\theta S_i\right)-1)
\end{align}
This yields the final closed-form expression for ${}^{\mathrm{SA}}\hat{G}_{aiaj}$ and ${}^{\mathrm{SA}}\hat{G}_{aibi}$,
\begin{align}
    \exp\left(\theta\ ^\text{SA}\hat{G}_{aiaj}\right) &= \hat{I} + \sum_{n=1}^2 \left(k_n^{(1)}\ ^\text{SA}\hat{G}_{aiaj}+k_n^{(3)}\ ^\text{SA}\hat{G}_{aiaj}^3\right)\sin\left(\theta S_n\right)\label{eq:case_aiaj}
    \nonumber\\
    &\quad+ \sum_{n=1}^2 \left(k_n^{(2)}\ ^\text{SA}\hat{G}_{aiaj}^2+k_n^{(4)}\ ^\text{SA}\hat{G}_{aiaj}^4\right)(\cos\left(\theta S_n\right)-1)~.
\end{align}

The closed-form expression for $\exp\left(\theta\ ^\text{SA}\hat{G}_{aibi}\right)$ is identical to that of $\exp\left(\theta\ ^\text{SA}\hat{G}_{aiaj}\right)$ but with ${}^{\mathrm{SA}}\hat{G}_{aiaj}$ replaced by ${}^{\mathrm{SA}}\hat{G}_{aibi}$.

\subsection{Case ${}^{\mathrm{SA}}\hat{G}_{aibj}$}

Now we turn to the anti-Hermitian version of Eq. (\ref{eq:sa_case4}), i.e.,\ the case with four unique indices coming from the intermediate singlet single connection. Here, the polynomial relation becomes
\begin{equation}
    ^\text{SA}\hat{G}_{aibj}^9 = A\ ^\text{SA}\hat{G}_{aibj}+B\ ^\text{SA}\hat{G}_{aibj}^3+C\ 
 ^\text{SA}\hat{G}_{aibj}^5+D\ ^\text{SA}\hat{G}_{aibj}^7
\end{equation}
with $A=-\frac{1}{4}$, $B=-\frac{15}{8}$, $C=-\frac{35}{8}$ and $D=-\frac{15}{4}$.
Following the same procedure as the previous cases gives the closed-form expression as
\begin{align}
    \exp\left(\theta\ {}^\text{SA}\hat{G}_{aibj}\right) &= \hat{I} + \sum_{n=1}^4\left(k_n^{(1)}\ {}^\text{SA}\hat{G}_{aibj} + k_n^{(3)}\ {}^\text{SA}\hat{G}_{aibj}^3\right.\label{eq:case_aibj}\\\nonumber
    &\quad\quad\quad\quad\left. + \ k_n^{(5)}\ {}^\text{SA}\hat{G}_{aibj}^5+k_n^{(7)}\ {}^\text{SA}\hat{G}_{aibj}^7\right)\sin\left(S_n\theta\right)\\\nonumber
    & \quad +\sum_{n=1}^4\left(k_n^{(2)}\ {}^\text{SA}\hat{G}_{aibj}^2 + k_n^{(4)}\ {}^\text{SA}\hat{G}_{aibj}^4\right.
    \\\nonumber&\quad\quad\quad\quad\left. + \ k_n^{(6)}\ {}^\text{SA}\hat{G}_{aibj}^6 + \ k_n^{(8)}\ {}^\text{SA}\hat{G}_{aibj}^8\right)\left(\cos\left(S_n\theta\right)-1\right)\nonumber
\end{align}
with the coefficients given in Table \ref{tab:case_aibj}.
\begin{table}[H]
    \centering
\caption{Values of coefficients in Eq. (\ref{eq:case_aibj}).}
\begin{tabular}{ c | c | c | c | c | c | c | c | c | c }
 $n$ & $k_n^{(1)}$ & $k_n^{(3)}$ & $k_n^{(5)}$ & $k_n^{(7)}$ & $k_n^{(2)}$ & $k_n^{(4)}$ & $k_n^{(6)}$ & $k_n^{(8)}$ & $S_n$\\
 \hline
1 & $\frac{2}{3}$          & $\frac{13}{3}$          & $\frac{22}{3}$          & $\frac{8}{3}$           & -$\frac{2}{3}$    & $-\frac{13}{3}$  & $-\frac{22}{3}$  & $-\frac{8}{3}$    & $1$ \\
2 & $-\frac{\sqrt{2}}{42}$ & $-\frac{\sqrt{2}}{6}$   & $-\frac{\sqrt{2}}{3}$   & $-\frac{4\sqrt{2}}{21}$ & $\frac{1}{42}$    & $\frac{1}{6}$    & $\frac{1}{3}$    & $\frac{4}{21}$    & $\sqrt{2}$ \\
3 & $-\frac{8\sqrt{2}}{3}$ & $-\frac{44\sqrt{2}}{3}$ & $-\frac{52\sqrt{2}}{3}$ & $-\frac{16\sqrt{2}}{3}$ & $\frac{16}{3}$    & $\frac{88}{3}$   & $\frac{104}{3}$  & $\frac{32}{3}$    & $\frac{\sqrt{2}}{2}$ \\
4 & $\frac{128}{21}$       & $\frac{64}{3}$          & $\frac{64}{3}$          & $\frac{128}{21}$        & $-\frac{256}{21}$ & $-\frac{128}{3}$ & $-\frac{128}{3}$ & $-\frac{256}{21}$ & $\frac{1}{2}$
\\\hline
\end{tabular}
\label{tab:case_aibj}
\end{table}

\subsection{Case ${}^\mathrm{SA}\hat{G}^{\prime}_{aibj}$}

The spin-adapted double that comes from the intermediate triplet excitation with four unique indices has the polynomial relation,

\begin{equation}
    ^\text{SA}\hat{G}_{aibj}^{\prime 11} = A\ ^\text{SA}\hat{G}_{aibj}^{\prime}+B\ ^\text{SA}\hat{G}_{aibj}^{\prime 3}+C\ 
 ^\text{SA}\hat{G}_{aibj}^{\prime 5}+D\ ^\text{SA}\hat{G}_{aibj}^{\prime 7} +E\ ^\text{SA}\hat{G}_{aibj}^{\prime 9}
\end{equation}
with $A=-\frac{113}{6}$, $B=-\frac{587}{6}$, $C=-\frac{613}{3}$, $D=-176$ and $E=-48$.
The closed-form expression is obtained using the same procedure as above,
\begin{align}
    \exp\left(\theta\ ^\text{SA}\hat{G}^{\prime}_{aibj}\right) &= \hat{I} + \sum_{n=1}^5\left(k_n^{(1)}\ ^\text{SA}\hat{G}^{\prime}_{aibj} + k_n^{(3)}\ ^\text{SA}\hat{G}_{aibj}^{\prime 3} + k_n^{(5)}\ ^\text{SA}\hat{G}_{aibj}^{\prime 5}\right.\label{eq:case_prime_aibj}\\\nonumber
    &\quad\quad\quad\quad\left. + \ k_n^{(7)}\ ^\text{SA}\hat{G}_{aibj}^{\prime 7} + k_n^{(9)}\ ^\text{SA}\hat{G}_{aibj}^{\prime 9}\right)\sin\left(S_n\theta\right)\\\nonumber
    &\quad + \sum_{n=1}^5\left(k_n^{(2)}\ ^\text{SA}\hat{G}_{aibj}^{\prime 2} + k_n^{(4)}\ ^\text{SA}\hat{G}_{aibj}^{\prime 4} + k_n^{(6)}\ ^\text{SA}\hat{G}_{aibj}^{\prime 6}\right.\\\nonumber
    &\quad\quad\quad\quad\left. + \ k_n^{(8)}\ ^\text{SA}\hat{G}_{aibj}^{\prime 8}+k_n^{(10)}\ ^\text{SA}\hat{G}_{aibj}^{\prime 10}\right)\left(\cos\left(S_n\theta\right)-1\right)\nonumber
\end{align}
with the coefficients given in Table \ref{tab:case_prime_aibj}.
\begin{table}[H]
    \centering
\caption{Values of coefficients in Eq. (\ref{eq:case_prime_aibj}).}
\begin{tabular}{ c | c | c | c | c | c | c | c | c | c | c | c }
 $n$ & $k_n^{(1)}$ & $k_n^{(3)}$ & $k_n^{(5)}$ & $k_n^{(7)}$ & $k_n^{(9)}$ & $k_n^{(2)}$ & $k_n^{(4)}$ & $k_n^{(6)}$ & $k_n^{(8)}$ & $k_{n}^{(10)}$ & $S_n$\\
 \hline
1 & $\frac{\sqrt{2}}{1150}$ & $\frac{11 \sqrt{2}}{690}$ & $\frac{133 \sqrt{2}}{1725}$ & $\frac{16 \sqrt{2}}{115}$ & $\frac{48 \sqrt{2}}{575}$ & $- \frac{1}{1150}$ & $- \frac{11}{690}$ & $- \frac{133}{1725}$ & $- \frac{16}{115}$ & $- \frac{48}{575}$ & $\sqrt{2}$\\
2 & $\frac{8 \sqrt{2}}{5}$ & $\frac{404 \sqrt{2}}{15}$ & $\frac{308 \sqrt{2}}{3}$ & $\frac{608 \sqrt{2}}{5}$ & $\frac{192 \sqrt{2}}{5}$ & $- \frac{16}{5}$ & $- \frac{808}{15}$ & $- \frac{616}{3}$ & $- \frac{1216}{5}$ & $- \frac{384}{5}$ & $\frac{\sqrt{2}}{2}$ \\
3 & $- \frac{54 \sqrt{3}}{25}$ & $- \frac{171 \sqrt{3}}{5}$ & $- \frac{2718 \sqrt{3}}{25}$ & $- \frac{576 \sqrt{3}}{5}$ & $- \frac{864 \sqrt{3}}{25}$ & $\frac{162}{25}$ & $\frac{513}{5}$ & $\frac{8154}{25}$ & $\frac{1728}{5}$ & $\frac{2592}{25}$ & $\frac{\sqrt{3}}{3}$ \\
4 & $- \frac{16 \sqrt{3}}{75}$ & $- \frac{56 \sqrt{3}}{15}$ & $- \frac{1192 \sqrt{3}}{75}$ & $- \frac{112 \sqrt{3}}{5}$ & $- \frac{192 \sqrt{3}}{25}$ & $\frac{32}{75}$ & $\frac{112}{15}$ & $\frac{2384}{75}$ & $\frac{224}{5}$ & $\frac{384}{25}$ & $\frac{\sqrt{3}}{2}$\\
5 & $\frac{432 \sqrt{3}}{115}$ & $\frac{2952 \sqrt{3}}{115}$ & $\frac{1368 \sqrt{3}}{23}$ & $\frac{6192 \sqrt{3}}{115}$ & $\frac{1728 \sqrt{3}}{115}$ & $- \frac{2592}{115}$ & $- \frac{17712}{115}$ & $- \frac{8208}{23}$ & $- \frac{37152}{115}$ & $- \frac{10368}{115}$ & $\frac{\sqrt{3}}{6}$
\\\hline
\end{tabular}
\label{tab:case_prime_aibj}
\end{table}

\subsection{Wave function parameter reduction}

The most prominent application of our finding is for the factorized unitary coupled cluster (fUCC) method. The closed form expressions derived in this work allow it to efficiently formulate for the spin-adapted fUCCSD, as well as efficiently formulate spin-adapted fermionic-ADAPT, which will then be guaranteed to converge to the correct spin symmetry.

As using spin-adapted operators is equivalent to being in the space of configuration state functions instead of the space of determinants, the number of parameters is reduced accordingly.
As an example, we here consider parameterization of a factorized unitary coupled cluster expansion to singles and doubles.
\begin{table}[H]
    \centering
\caption{Number of variational parameters using factorized unitary coupled cluster singles doubles and spin-adapted factorized unitary coupled cluster singles doubles for different system sizes.}
\begin{tabular}{ c | c | c | c | c | c | c | c | c}
  & (2,2) & (4,4) & (6,6) & (8,8) & (10,10) & (12,12) & (14,14) & (16,16)\\
 \hline
fUCCSD & 3 & 26 & 117 & 360 & 875 & 1818 & 3381 & 5792\\
SA-fUCCSD & 2 & 14 & 54 & 152 & 350 & 702 & 1274 & 2144
\\\hline
\end{tabular}
\label{tab:ucc_params}
\end{table}
In Table \ref{tab:ucc_params}, the number of variational parameters for different sizes for fUCCSD and SA-fUCCSD expansions can be seen.
The reduction of the number of parameters becomes about 60\% for the larger systems and goes towards a 66\% reduction asymptotically.
Another direct benefit of using spin-adapted operators is that the number of operators to measure in an iteration of a fermionic-ADAPT\cite{Grimsley2019-ix} will also be reduced, equivalent to the parameter reduction for a fUCCSD.

\section{Conclusion}

In this work, exact closed-form expressions have been derived for unitary spin-adapted fermionic double excitation operators.
These expressions can be utilized for an efficient implementation of these operators in unitary product state codes that target conventional hardware.
Moreover, our findings might also provide guidance about how to construct circuits for spin-adapted double excitation operators on quantum hardware. This will be part of future research.

\acknowledgments
We acknowledge the support of the Novo Nordisk Foundation (NNF) for the focused research project ``Hybrid Quantum Chemistry on Hybrid Quantum Computers'' (grant number  NNFSA220080996).

\section*{DATA AVAILABILITY}
The data that support the findings of this study are available from the corresponding author upon reasonable request.

\newpage
\bibliographystyle{unsrt}
\bibliography{literature}

\end{document}